\documentclass[useAMS,usenatbib]{mn2e}

\usepackage{graphicx}
\usepackage{float}
\usepackage{verbatim}
\usepackage{color}
\usepackage{amsmath}
\usepackage{hyperref}
\usepackage{url}

\usepackage{mymacros}	
\usepackage[font=small]{caption}

\usepackage{graphicx}%rotating, graphicx,dsfont}
\usepackage{xcolor}
\usepackage{amssymb, amsmath}
\usepackage[export]{adjustbox}
\hypersetup{colorlinks=true,citecolor=blue}
\usepackage[utf8]{inputenc}
\usepackage{amsmath, amssymb, bm}
\usepackage{multirow}
\renewcommand{\vec}[1]{\bmath{#1}}

\newcommand{\bea}{\begin{eqnarray}}
\newcommand{\eea}{\end{eqnarray}}
\newcommand{\bdm}{\begin{displaymath}}
\newcommand{\edm}{\end{displaymath}}

\newcommand{\eq}[1]{Eq.~(\ref{#1})}

\newcommand{\fig}[1]{Figure~\ref{#1}}

\def\ie{{\em i.e.}~}

\def\Mpc{\, h^{-1} \, {\rm Mpc}}

\def\aap{Astronomy \& Astrophysics}
\def\apj{The Astrophysical Journal}
\def\mnras{Monthly Notices of the Royal Astronomical Society}

\title[Lagrangian Bias]{Halo bias in Lagrangian Space: \\ Estimators and theoretical predictions }

\author[Modi et al.]{Chirag Modi$^{1,2}$\thanks{E-mail: modichirag@berkeley.edu}, Emanuele Castorina$^{1}$ and Uro\v s Seljak$^{1,2}$\\
$^{1}$Berkeley Center for Cosmological Physics Campbell Hall 341, University of California, Berkeley CA 94720\\
$^{2}$Department of Physics, University of California, Berkeley CA 94720}
\begin{document}

\date{Received ; in original form }

\pagerange{\pageref{firstpage}--\pageref{lastpage}} \pubyear{2014}

\maketitle

\label{firstpage}

\begin{abstract}
We present several methods to accurately estimate Lagrangian bias parameters and substantiate them using simulations. In particular, we focus on the quadratic terms, both the local and the non local ones, and show the first clear evidence for the latter in the simulations. Using Fourier space correlations, we also show for the first time, the scale dependence of the quadratic and non-local bias coefficients. For the linear bias, we fit for the scale dependence and demonstrate the validity of a consistency relation between linear bias parameters. 
Furthermore we employ real space estimators, using both cross-correlations and the Peak-Background Split argument. This is the first time the latter is used to measure anisotropic bias coefficients. We find good agreement for all the parameters among these different methods, and also good agreement for local bias with ESP$\tau$ theory predictions. We also try to exploit possible relations among the different bias parameters.
Finally, we show how including higher order bias reduces the magnitude and scale dependence of stochasticity of the halo field. 
\end{abstract}

\begin{keywords}
cosmology: theory, large-scale structure of Universe – methods: analytical, numerical
\end{keywords}

\section{Introduction}

The distribution of Large scale structures (LSS) is one of the most important tools to study our Universe. With the recently concluded and upcoming LSS surveys such as BOSS\footnote{\url{https://www.sdss3.org/index.php}}, DESI\footnote{\url{http://desi.lbl.gov/}}, Euclid\footnote{\url{http://www.euclid-ec.org/}}, LSST\footnote{\url{https://www.lsst.org/}} etc. the vast amount of data will provide enough statistical power that errors in the analysis will be dominated by poor theoretical understanding of LSS. These surveys generally observe tracers such as galaxies, which unlike in weak lensing studies, do not perfectly trace the underlying matter density. Making this connection is a two step process- to realize that galaxies reside in collapsed dark matter halos following some statistical distribution- and that these halos themselves are biased tracers of matter distribution. In this paper, we focus on the latter of these and study novel techniques to measure halo bias. For a recent review on galaxy bias see \cite{BiasRev}.

In full generality the relation between the halo density field and the dark matter contains information about all processes relevant for the formation of halos, as well as stochastic contributions arising due to the fact that halos come in a finite number.
However on large enough scales, we can hope to treat the problem perturbatively, and characterize the bias relation with a relatively small number of free parameters we can marginalize over.
The simplest way to get physical intuition of halo bias comes from the so-called Peak-Background split argument (PBS) \citep{Kaiser1984,Kaiser1989,Mo1996,Sheth1999}. In the original formulation of PBS, halos are regions where the value of the dark matter density field exceeds some critical threshold. This threshold can be crossed more easily in presence of a positive large scale dark matter fluctuations, and conversely less easily for a negative one. This means that, on average, overdense regions host more halos than the mean. 
Linear bias, ($b_1$),  can therefore be defined as the linear response of the halo overdensity field ($\delta_h$), to the presence of long wavelength perturbations ($\delta_m$), 
\be
\delta_h = b_1 \delta_m
\ee
The above relation can be generalized to any order in the density field, leading to the well known result by \cite{Fry1993},
\be
\label{eq:FryGatz}
1+\delta_h(x) = \sum_{n = 0}^\infty \frac{b_n}{n!}\delta_m^n(x)\,,
\ee
where however one must also enforce integral constraint by imposing that $\langle\delta_h\rangle = 0$.

Typically the bias expansion is written in two ways: either by relating the the halo field to the dark matter field at time the observations are made, \ie at low redshift, or by identifying in the initial Gaussian field regions which are more likely to collapse into halos at a later time. 
The former is called a Eulerian bias approach, the latter a Lagrangian bias approach.
While the two approaches can be shown to be mathematically equivalent, each method has its own pros and cons and in this work, we use our simulations to focus on the latter. 

It has been recently shown that the statistics of the halos does not only depend on the dark matter density field, as in \eq{eq:FryGatz}, but also on spatial derivatives of the density field,\citep{Desjacques2010, Musso2012,Baldauf2015} and on the tidal fields \citep{,Chan2012,Baldauf2012,Saito2014}, each of which with its own new bias coefficient. 
These new terms, with abuse of notation, have been called non local bias coefficients, since they contain gradients of the the density field or of the gravitational potential. They have been shown to be important using simulations in the Eulerian approach \citep{Chan2012,Baldauf2012,Saito2014}, as they can be generated by non linear gravitational evolution, and can actually be predicted in Perturbation Theory (PT).
 As a matter of fact a complete basis would include all possible fields one can construct at any given order in perturbation theory compatible with the symmetry of the problem \citep{McDonald2009,Assassi2014,Senatore2015,Mirbabayi2015,Fujita2016,Vlah2016}.

The relevance of non local terms in the Lagrangian approach is still under debate. Using N-body simulations, works such as \cite{Desjacques2010, Paranjape2013, Biagetti2014, Baldauf2015,Castorina16} do clearly show the presence of a scale dependent bias term at linear level and its measured value agrees fairly well with analytical models of structure formation \citep{Paranjape2013,Castorina16}. 

At second order the leading non local term is proportional to the traceless shear $s^2$, also called tidal field, defined as
\be
\label{eq:s2}
s^2({\bx}) \equiv \sum_{ij} s^2_{ij}(\bx) \;\text{, }\; s^2_{ij}(\bx) = (\frac{\partial_i \partial_j}{\partial^2} - \frac{1}{3}\delta_{ij}^K) \delta(\bx)\;,
\ee
where the indices ${i,j}$ run over all the three directions and $\delta(\bx)$ is the dark matter overdensity. This new term has been measured in Eulerian space using simulations by \cite{Chan2012,Baldauf2012,Saito2014,Hoffmann2015}, and it is nowadays included in all analysis of galaxy redshift surveys \citep{Gil-Marin2015, Beutler2016, Sanchez2017}. However, it is important to compare the measured Eulerian tidal bias against both of its origins - the non-zero tidal bias that is expected from non-linear gravitational evolution as well as the initial Lagrangian tidal bias, since an incorrect assumption on the latter could affect the determination of cosmological parameters. We touch upon this in our work. 

Tidal bias in the Lagrangian picture is non-zero if the shear is sampled around proto-halos differently than around random positions.  We actually do expect the shear to play an important role in the process of halo collapse, that is very likely to happen in an ellipsoidal fashion \citep{BondMyers96, Sheth1999, Desjacques2008}. The work in \cite{Sheth2013,Castorina16} have not shown clear evidence for tidal bias in Lagrangian space using N-body simulations. It is therefore interesting, and it is one of our main goals, to check whether this bias coefficient has non-zero value in a Lagrangian scheme and what would be the implications.

In addition to finding evidence of new bias coefficients, there is also another good reason to study Lagrangian halo bias. In Lagrangian Space it is more natural to write down relation among the different parameters, since, contrary to the Eulerian picture, non-linear evolution and the biasing scheme are decoupled from each other. For similar reasons, the evolution of bias with redshift, an important issue for galaxy surveys that spans a wide redshift range, is better understood in Lagrangian space.
Finally, analytic approaches to predict the value of bias coefficients are built in the Lagrangian formalism, and can be used to put priors on some of the bias parameters. As we are ultimately interested in cosmological analysis of LSS data, a combination of priors and relations between the bias coefficients would be of great help to narrow the error bars on cosmological parameters.

Recently \cite{Vlah2016} have shown that at 1-loop in Lagrangian Effective Perturbation theory, the measurements in redshift space of the halo multipoles are very well described if one adopts the following bias expansion,

\begin{equation}
\label{eq:real-bias}
\delta_h(\bx) = b_1 \delta(\bx) + \frac{b_2}{2}[\delta^2(\bx) - \langle \delta^2(\bx)\rangle] + b_{s^2}[s^2(\bx) - \langle s^2(\bx) \rangle]\;,
\end{equation}
\ie stopping at second order and assuming that third order terms are only generated by gravitational evolution. 
Our goal is to estimate the three bias coefficients in the above equation in N-body simulations using different techniques, and to compare them to analytic predictions. We are primarily interested in scale dependent bias, linear and non linear, and in the first clear evidence of tidal shear in the bias expansions.
We will make use of three different estimators: cross correlations in Fourier Space, cross correlations in Real space, and direct implementation of the PBS argument to simulations.

We organize the paper as follows. We start by describing our suite of simulations in Section \ref{Sec:sim} followed by Section \ref{Sec:bF} where we describe our approach to estimate bias using cross power spectra in Fourier space, showing results for the bias coefficients up to second order. In Section \ref{Sec:bReal}, estimators in real space that make use of the PDF of the density field are discussed, in particular focusing on their agreement with the Fourier space measurements. In Section \ref{Sec:bPBS} we present a direct application of PBS to simulations that allows to recover the value of the scale independent part of the bias coefficients. After establishing the agreement between different estimators, in Section \ref{Sec:rel}, we present various relations amongst different parameters. Section \ref{Sec:stoch} deals with the halo-bias stochasticity, and how much it is reduced by introducing a more complicated bias model. Finally, 
section \ref{Sec:conclude} summarizes the main results, drawing conclusions and future prospects. 

{\color{black} Since we work in Lagrangian space, throughout the text, we will refer to the proto-halos as dark matter halo  \ie we identify the collection of the particles in the simulation which forms the halo at the final redshift, and this collection of particles mapped to the initial redshift forms our (proto)halo in the Lagrangian space.  This mapping is easily achieved by storing the particles ID at each snapshot. Then, to define the proto-halo field, we calculate the center-of-mass of these particles using their positions at the initial conditions and use them as (proto)halo positions.}

\section{Simulations}
\label{Sec:sim}

The simulations we used for the analysis have been produced using the FastPM\footnote{\url{https://github.com/rainwoodman/fastpm}} code \citep{Feng2016}. FastPM is a tool to generate non linear dark matter and halo fields in a quick way, employing a number of approximations to reproduce the results of full N-body simulation. Despite its approximate nature, in \cite{Feng2016} it was shown that the code performs extremely well on various benchmarks such as dark matter power spectrum, the halo mass function and halo power spectrum. {\color{black} They also recommend using the halo catalogs after abundance matching them, and hence we abundance match our catalogs with MICE mass function \citep{Crocce2010} after correcting for systematic error in number of halo particles caused by FOF method as given in \cite{Warren2006}.} We refer the reader to \cite{Feng2016} for more details of the tests performed.

{\color{black}The simulations have a flat $\Lambda$CDM cosmology model with the Hubble parameter $h =0.6711$, matter density $\Omega_M = 0.3175$, baryon density $\Omega_b = 0.049$ and spectral index $n_s = 0.9624$. The linear power spectrum used was generated with CAMB. We evolved $2048^3$ particles in periodic cubic boxes of size 690, 1380 and 3000 $\Mpc$, giving the particle mass of $3.37 \times 10^9 M_{\odot}/h$, $2.696 \times 10^{10} M_{\odot}/h$ and $2.77 \times 10^{11} M_{\odot}/h$ respectively. The initial conditions of particle displacement and velocity were computed at 2nd order in Lagrangian Perturbation Theory at redshift $z=39$. Particles were then evolved to redshift $z=0$ in 40 timesteps linearly spaced in scale factor.} The halos were identified using FOF halo finder in NbodyKit\footnote{\url{https://github.com/bccp/nbodykit}} with the optimization presented in \cite{Feng2016a} using a linking length of 0.2. The smallest halo identified in every simulation consists of a 100 particles. We ran 5 independent realizations for every box and all the results presented are average of these runs. Unless stated otherwise, we show results at a single redshift, $z=0$ \ie the collections of particles that make up the (proto)halo field in Lagrangian space were made by identifying the halos in the simulation snapshot at $z=0$.

To calculate power spectra, we interpolate the halos on $512^3$ grid using Cloud-In-Cell (CIC) interpolation and deconvolve the CIC window. To keep the comparison consistent, simulation snapshots were gridded on the same $512^3$ grid to get dark matter density field and corresponding spectra.

\section{Bias estimation in Fourier space}
\label{Sec:bF}
\begin{figure}
\includegraphics[width=\columnwidth]{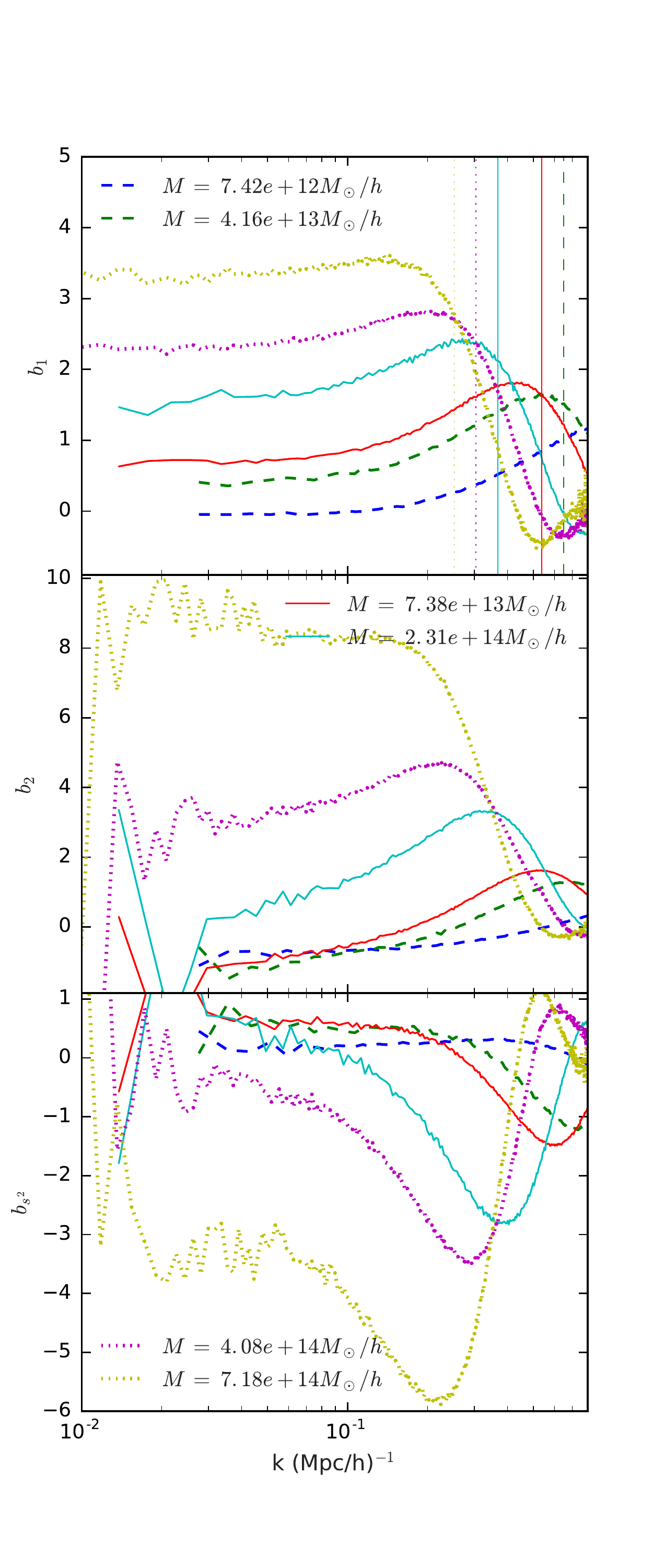}
\vspace*{-10mm}
\caption{{\bf Bias parameters as function of $k$}: We show $b_1$, $b_2$ and $b_{s^2}$ as measured from Fourier space estimator. For clarity, we only show 2 mass bins from every box and different line-styles (dashed, solid and dotted) correspond to different boxes (of size $ L = 690, 1380, 3000\ Mpc/h$ respectively), from which the corresponding mass bins (specified by color) are picked. The dependence on the wavenumber is the shared by all the parameters: constant piece on large scales followed by $k^2$-like piece on intermediate scales followed by a cutoff around the halo scale. This scale is shown with vertical lines in top-panel in corresponding colors for different masses.}
\label{fig:bias-kgrid}
\end{figure}

The easiest thing to measure in Fourier Space is linear bias. One way to estimate it is simply taking the ratio of halo-matter cross spectrum with the matter-auto spectrum.
\be
b_1 =\frac{P_{h\delta}}{P_{\delta \delta}} 
\ee
where in our notation the power spectrum between the $X$ and $Y$ field reads $P_{XY}$. Throughout the paper, we refer to the halo-field with subscript $(h)$ while ($\delta$) will refer to the Lagrangian overdensity field scaled to z=0 using linear theory.
In \fig{fig:bias-kgrid}, top panel,  we show linear bias as measured from our FastPM runs for several halo mass bins. The common features shared by all the halos are: a constant piece on large scales, and then a $k^2$-like piece on intermediate scales followed by a cut off, approximately at the scale of the halo. Similar plots for linear bias can be found for instance in \cite{Baldauf2015}. 
A convenient parametrization of linear bias is therefore \citep{Musso2012,Chan2015,Castorina16}
\be
\label{eq:b1k}
b_1(k) = \left(b_{10}+2 b_{11}\frac{\mathrm{d} \,\log W(k R)}{\mathrm{d}\, \log S}\right)W(k R)
\ee
where $S\equiv \sigma^2$ is the variance of the linear field smoothed on the scale of the halo $R$, $S \equiv (2\pi)^{-3}\int \mathrm{d}^3k \,P(k) W(kR)^2$.
For instance, for a Gaussian windows, $W(kR) = e^{- k^2 R^2}$, the scale dependent term $b_{11}$ is exactly proportional to $k^2$.
In this form the two bias coefficients $b_{10}$ and $b_{11}$ are both dimensionless.
Since neither the halo scale, $R$, nor the average halo shape, $W(k R)$, are known, fitting the above equation to the simulations results requires some care. {\color{black} In principle the halo scale and the window function should be considered as part of the model we are trying to test, see for instance \cite{Chan2015,Chan16}.} To be conservative, and for comparison with previous work, we assume proto-halos of mass $M$ are spherical patches of size $R= (3 M / 4 \pi\rho_m )^{1/3}$, such that the window function $W(k R)$ is top-hat of radius $R$ in real space\footnote{Another choice could be a Gaussian window, $W_G=e^{-k^2R^2/2}$, such as employed by  \cite{Baldauf2015}, but the fact that linear bias in \fig{fig:bias-kgrid} crosses zero at large $k$ excludes this}. As a further safeguard, we fit for $b_1(k)$ only upto the first peak, after which the halo window function starts to dominate over the scale dependence of bias. {\color{black} More details can be found at the end of this section where our fitting procedure is described.}

\begin{figure}
\includegraphics[width=\columnwidth]{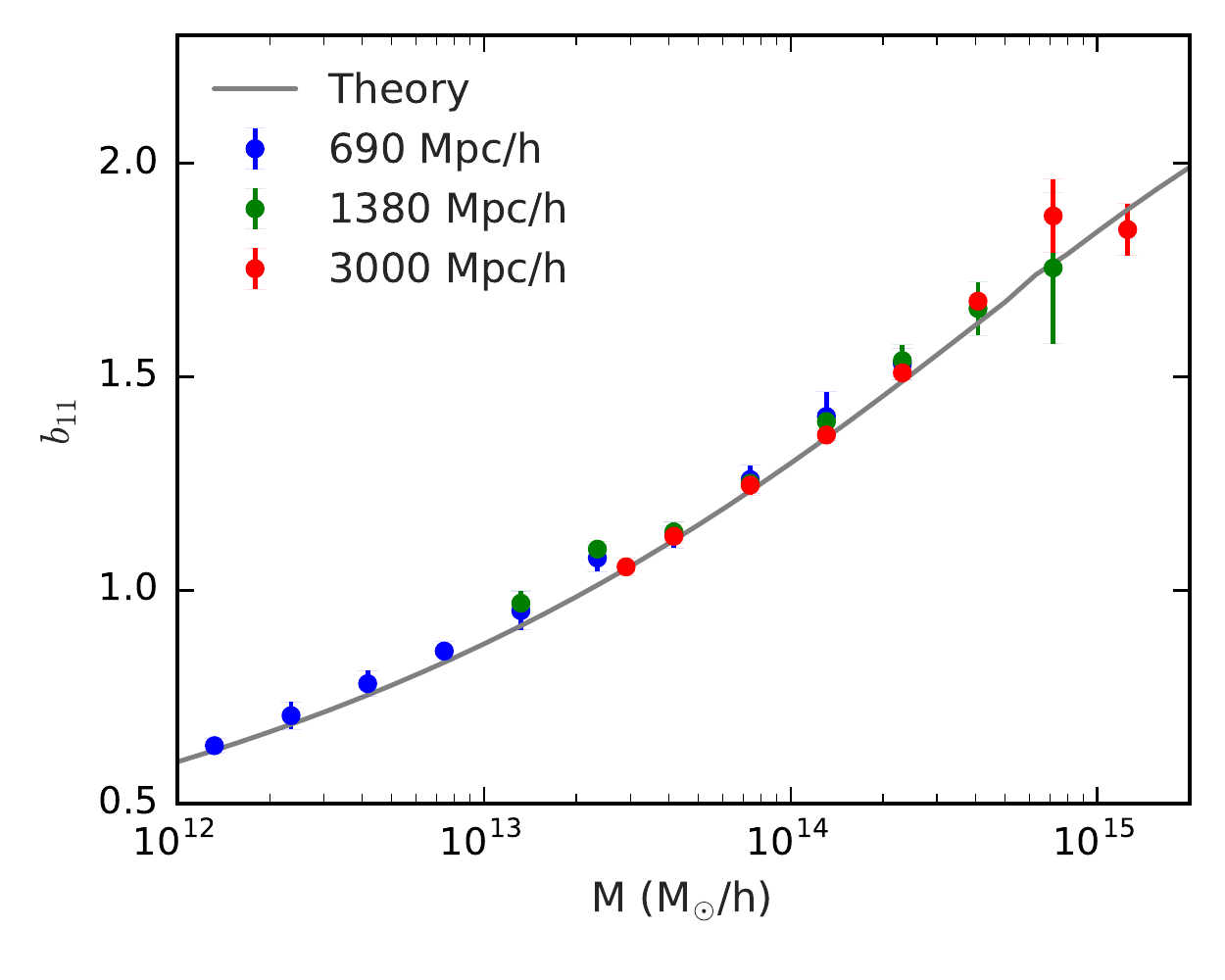}
\caption{{\bf $b_{11}$ as a function of mass:} The values are estimated from the best fit values for $b_{1}$ and using \eq{eq:b1k}, assuming a tophat window $W(kR)$ at the halo scale. We will also follow the scheme of the colors blue, green and red (B, G, R) representing different box sizes throughout this paper unless explicitly mentioned otherwise, or when the box size is not important.}
\label{fig-b11_mass}
\end{figure}

At this time we can fit for the scale independent piece of linear bias, $b_{10}$, as well as for the scale dependent one, $b_{11}$, in \eq{eq:b1k}. A comparison of $b_{10}$ with theoretical models and with other estimators for linear bias on large scales will be done in the next section, whereas here we only show, in Figure \ref{fig-b11_mass}, results for $b_{11}$ as a function of mass, and how it compares to the analytic prediction shown as the continuous line. \\
{\color{black} Since measuring the non-local shear bias is one of the primary motivations of this paper, we compare our numbers against analytic predictions from a theory that includes the effects of the shear field on halo formation and hence predicts a non-zero shear bias. We use the $\text{ESP}\tau$ model from (\cite{Castorina16}) which is an extension of Peaks theory (BBKS)(\cite{Bardeen1986}) that includes the Excursion set constraints (ESP) \citep{MS2012,Paranjape2013} and a treatment of ellipsoidal collapse inspired by \cite{BondMyers96,Sheth2001}. The $\text{ESP}\tau$ model explicitly attempts to include the dependence of initial shear ($s^2$) field in determining the collapse of halos by modifying the critical over-density threshold ($B$) from its standard spherical collapse value of $\delta_c = 1.686$ to the following expression
\be 
\label{eq:ESPtau}
\delta_R \ge B = \delta_c + \beta \sqrt{s^2(R)}
\ee 
Since the tidal field is positive definite, in the Esp$\tau$ model halos have to be denser than the spherical case in order to collapse at a given redshift. The barrier $B$ explicitly depends on the smoothing scale, and therefore different halos will have different collapse density.
There is only one free parameter, $\beta$, entering the definition of the critical threshold, and it measures the relative importance of shear fluctuations to density fluctuations at the (proto)halo scale. It can have arbitrary dependence on cosmology and redshift, and testing these possible dependences is one of the goals of this work.
A first comparison of $\text{ESP}\tau$ with N-body simulations as well as explicit mathematical details on the model can be found in \cite{Castorina16}. 

We fit $\beta$ to the measurements at $z=0$, finding that $\beta=0.3$ provides a good fit to the data. Any degradation of the goodness of fit to the data at higher redshift would imply redshift dependence of our free parameter.}

$\text{ESP}\tau$ performs remarkably, if not surprisingly, well in comparison to the scale dependent piece $b_{11}$ over a $3$ orders of magnitude in mass.
Since the value of scale dependent bias is sensitive to the choice of the filter (see \eq{eq:b1k}), the agreement between the theory and the data suggests that defining proto-halos with a Top-Hat filter is a very good approximation. This finding is in contradiction with \cite{Chan2015}, and it deserves investigation in a future work.
As we will see later in Section \ref{Sec:rel} the performance of $\text{ESP}\tau$ on scale dependent bias is important, as the value of $b_{10}$ and $b_{11}$ are related by mass conservation.

We would like to extend the results we have shown for $b_1$ to higher order bias coefficients, in particular we seek for a simple estimator of $b_{s^2}$.
First, notice that, since the bias relation in \eq{eq:real-bias} is written in real space, any estimators of second order bias coefficients in Fourier space will make use of convolutions, for instance the squared density field is trivially
\be
\delta^2(x) = \int \frac{\mathrm{d}^3k}{(2 \pi)^3} \,e^{ - i \vec{k} \cdot \vec{x}} \,\int \mathrm{d}^3q \,\delta(q)\delta(k-q)
\ee
and an analogous expression holds for $s^2$. 
We also recall that since the dark matter field is linear in Lagrangian space, cross correlations between the halos and a generic quadratic fields are very easy to write down, as any expectation value involving three fields vanishes. Within the model in \eq{eq:real-bias} we arrive to
\begin{align}
\label{eq:b2k-bs2k}
P_{h\delta^2}(k) &= \frac{b_2(k)}{2} P_{\delta^2 \delta^2}(k) + b_{s^2}(k) P_{s^2 \delta^2}(k)  \\
P_{h s^2}(k)  &=\frac{b_2(k)}{2} P_{s^2 \delta^2}(k) + b_{s^2}(k) P_{s^2 s^2}(k) \nonumber
\end{align}
where for instance 
\begin{eqnarray}
P_{\delta^2 \delta^2}(k) &=& \frac{2 k^3}{4\pi^2} \int_{r =0}^{\infty} dr \biggl( \ r^2 P_{\delta \delta}(kr) \times \\
	&& \int_{x=-1}^1 dx\  P_{\delta \delta}(k\sqrt{1 + r^2 -2rx}) \biggr) \nonumber
\end{eqnarray}
and similar expression can be written for $ P_{s^2 \delta^2}(k) $ and $ P_{s^2 s^2}(k)$. However, to be consistent in our approach, we do not evaluate these integrals numerically to estimate bias but evaluate these power-spectra directly from the corresponding simulations we find the halos in.\\
The linear system of equations in \eq{eq:b2k-bs2k} can be solved to obtain a value of $b_2$ and $b_{s^2}$ at each mode $k$ and for different halo populations. (To assist in solving, one can also re-write these equations in terms of $g^2 = s^2 - \frac{2}{3} \delta^2$ to decouple $s^2$ from $\delta^2$, though it is not necessary.) The solution is shown in the two bottom panels of \fig{fig:bias-kgrid}, which makes clear that the behavior as a function of $k$ we have seen for linear bias also applies to higher order bias coefficients.
Scale dependence of non linear bias would have, for instance, to be taken into account for Lagrangian Perturbation theory calculations beyond $1$-loop order. 

The bottom panel clearly shows that tidal bias is non zero for a variety of halo populations. This has important consequences for cosmological analyses using Lagrangian perturbation theory plus Lagrangian bias (\cite{Matsubara2008, Carlson2013, Wang2014, White2015, Vlah2016}).
As already known, $b_2$ is negative for low mass halos and then becomes large and positive for very massive halos. The opposite trend is observed for $b_{s^2}$, which is negative at the high mass end. This is expected, since shear is acting against the formation of the halos, whereas density enhances it. We emphasize that had we started with simply a local bias expansion at second order, we would have gotten different values for $b_2(k)$, in contradiction with the results of the next sections, and as further discussed at the end of Section \ref{Sec:bReal}.

The scale independent bias coefficients, such as those in \eq{eq:real-bias}, need to be extracted from the large scale values of any bias parameter $b(k)$ in \fig{fig:bias-kgrid}. {\color{black} To do this, we fit for functional form $b(k) = b_0 + b_k k^2$ for some constant $b_k$ on large scales ($k < 0.2$ h/Mpc) weighted by inverse variance of individual $b(k)$ data points (variance is calculated from the 5 simulations we have for each box size). Then $b_0$ corresponds to the scale independent large scale bias we are interested in. As discussed before, though the exact functional form to be fitted for should depend on the bias parameter being fit for as well as the window ($k^2$ corresponds to Gaussian window instead of the Tophat which we have assumed), we are only interested in the constant piece here and its value should be insensitive to the choice of the window on such large scales. Nevertheless, we confirm this with another fitting procedure- on large scales, \ie we fit with a constant value ($b_0$) using weighted least squares method as function of increasing $k_{max}$ using inverse-variance weights. The best fit value for $b_0$ then corresponds to the largest $k_{max}$ at which the sum of residuals per degree of freedom ($\sum_{k_i}^{k_{max}}(b(k) - b_0)^2 /N$) is minimum. The estimated values by both the procedure are consistent with each other and this reinforces the assumption that the best fit parameters for large scale biases are not sensitive to the choice of window function.}

While we do find clear evidence for non-zero tidal bias, getting good fits and precise value for it is hard due to greater oscillations on large scale, as seen in \fig{fig:bias-kgrid}. This is due to the combined effect of cosmic variance as well as degeneracy between $\delta^2$ and $s^2$ on large scales, making things noisy in the regime where the scale independent piece of bias should dominate, and it becomes more severe for our smallest simulation box. This also comes into play when we try to extend this method to $b_3$ (see Appendix \ref{Sec:Appendix-A}). Furthermore, unlike $b_1$, the functional form of the scale dependence of $b_{s^2}$ is highly non-trivial \citep{Matsubara2008,Musso2012,Castorina16}, and hence much more sensitive to the halo window, which is not exactly known, as well as $k_{max}$, up to which any such a fit is done. 

\section{Real Space bias estimators}
\label{Sec:bReal}
Cross correlations between the halos and other fields can also be written in real space. The advantage of working in Lagrangian Space is that we know the PDF of the density field, and therefore estimators for the cross correlation and for the bias coefficients can be written down without expensive pair counting. Following \citep{Szalay1988, Musso2012,Paranjape2013a,Biagetti2014,Castorina16a}, if $\Delta_0$ is the initial (Lagrangian) dark matter over-density scaled to $z=0$ by linear theory in a sphere of radius $R_0$ around halos of given mass $M$, for the $n$-th order bias coefficient we can write
\be
\label{eq:bn-real}
\hat{b}_n = \frac{S_{\times}^n}{S_0^{n/2}}\frac{1}{N}\sum_{i=1}^N H_n\left(\frac{\Delta_{0i}}{\sqrt{S_0}}\right)
\ee
  where the sum runs over all the $N$ halos falling into the mass bin, $H_n(x)$ is a probabilist's Hermite polynomial and {\color{black} the $\hat{}$ in $\hat{b}_n$ emphasizes that it is an estimator for $b_n$}. The rationale behind this estimator is that Hermite Polynomials are orthogonal polynomials of Gaussian random fields, therefore they will pick up only the $n$-th order in the density field contribution to the bias expansion. In the above equation $S_0 = \sigma^2(R_0)$ is the variance of the field on large scale, and $S_{\times} = \sigma^2(R,R_0)$ is the cross variance between the halo scale and the large scale.

\begin{figure}
\includegraphics[width=\columnwidth]{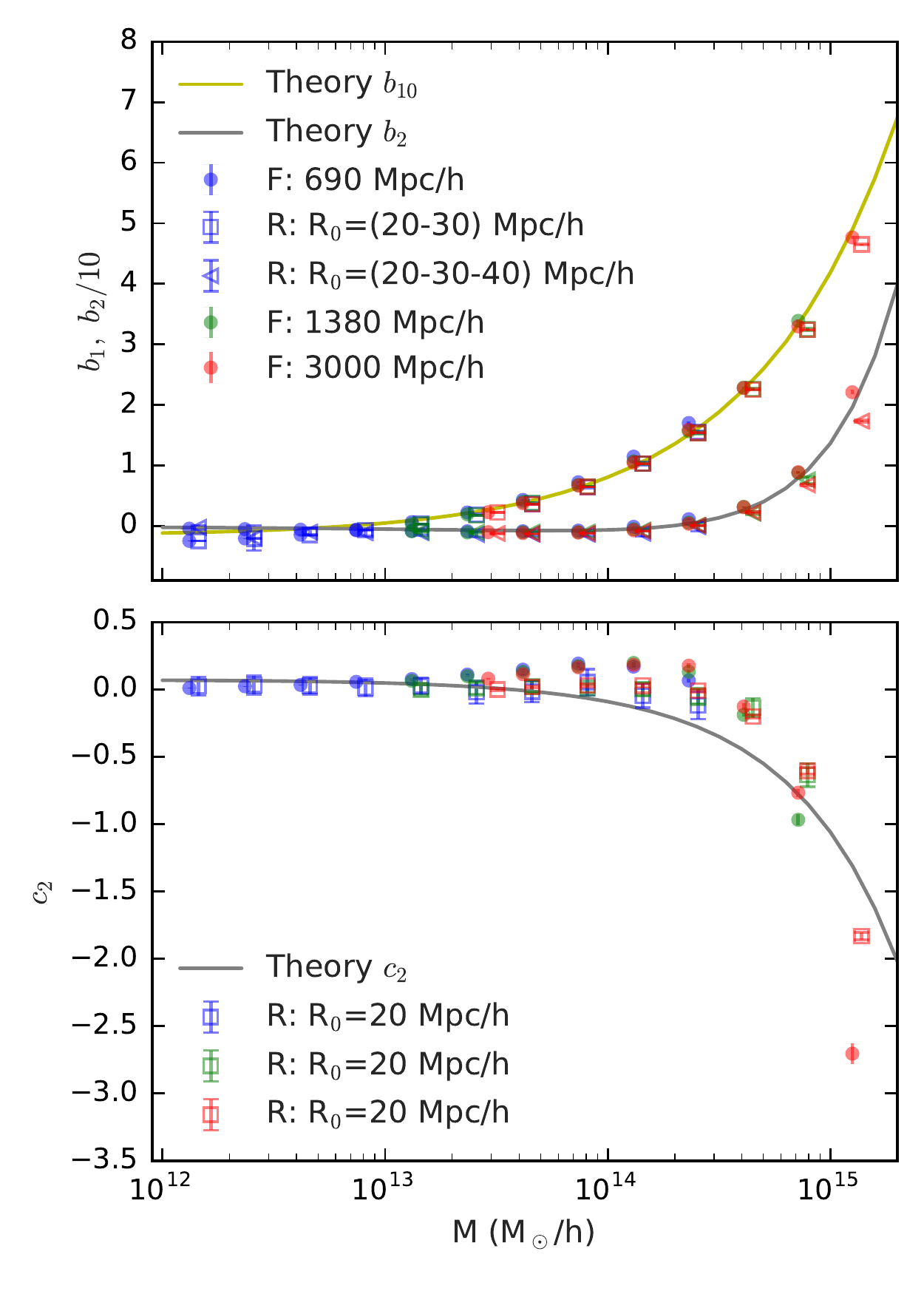}
\vspace*{-5mm}
\caption{{\bf Bias estimators $b_1$, $b_2$ and $b_{s^2}$ as function of mass}: agreement between real space, Fourier space estimator and $\text{ESP}\tau$ theory. For real space, we mention the large scale used to calculate the bias parameters. We use \eq{eq:b1real} and \eq{eq:b2real} for $b_{10}$ and $b_{20}$ respectively, and the smoothing scales used to extract scale independent part are mentioned. Real space points have been shifted along x-direction for clarity.}
\label{fig-b1M}
\end{figure}

\begin{figure}
 \includegraphics[width=\columnwidth]{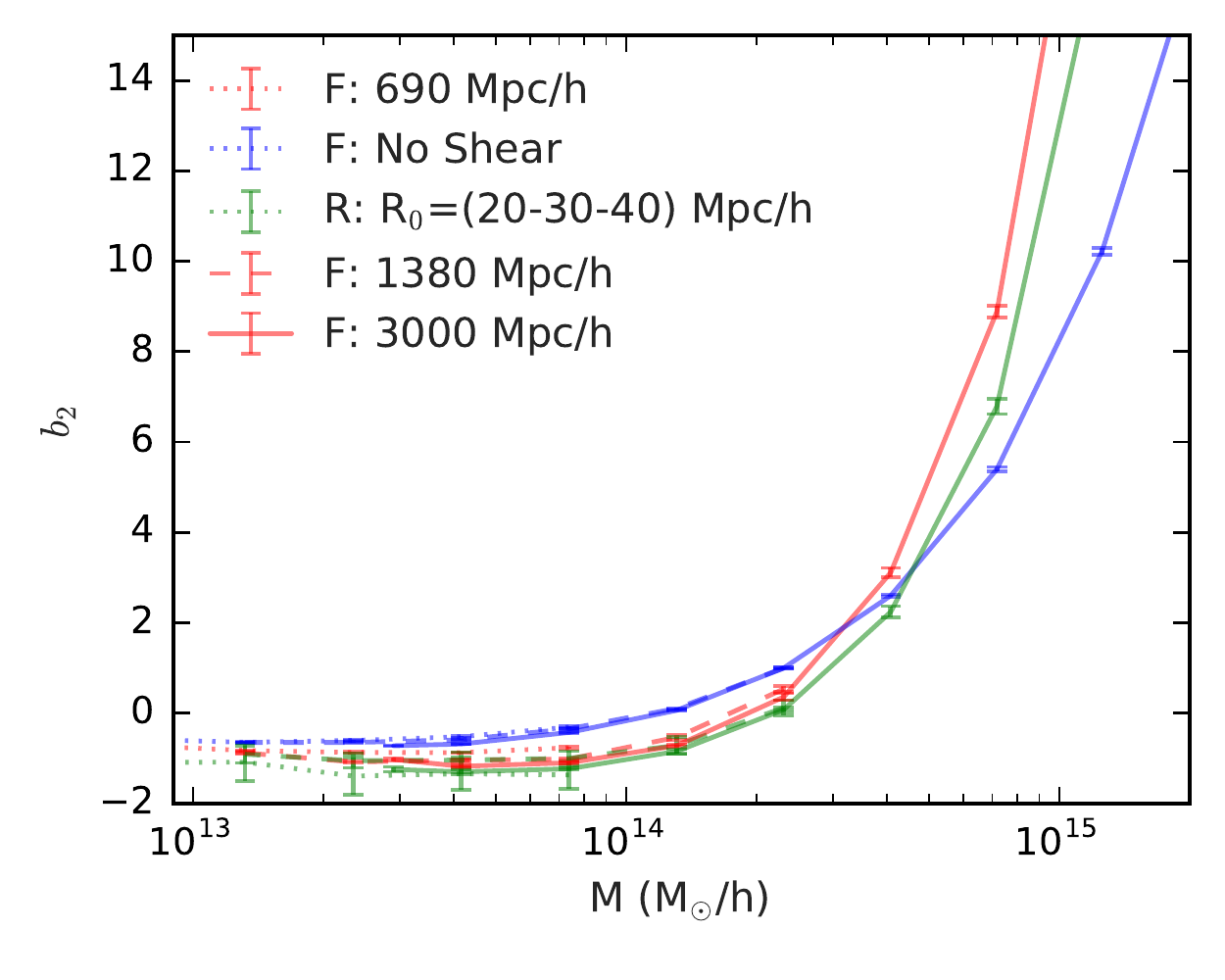}
 \caption{{\bf $b_2(M)$ without $b_{s^2}$}: Fourier space estimator for $b_2$ with (red points) and without (blue) including the shear in \eq{eq:b2k-bs2k} for 3 Gpc/h box. For comparison, we show the real space estimates (green boxed, shifted along x-axis for clarity) which are independent of the shear by construction. The two estimators are in agreement only if a non-vanishing tidal bias in Fourier space is considered. Though not shown here, the fit for $b_2$ without shear is also much worse for other boxes as well as PBS estimator, especially for high mass halos. Overall consistency with real space estimates is presented in Figure \ref{fig-b1M}.}
\label{fig-b2nos2}
\end{figure}

In real space, the convenient separation of scales we have seen in Fourier space is lost and the {\color{black} estimator in \eq{eq:bn-real} has contributions from both the scale independent and scale dependent components of the bias.} By Fourier transforming \eq{eq:b1k}, it is easy to see that measurements in real space can be fitted by
\be
\label{eq:b1real}
b_1 = b_{10}+ \epsilon_{\times} b_{11}
\ee
where $\epsilon_{\times} = d\,\log S_{\times} / d\,\log S_{0}$.
The scale independent piece can be extracted by using a very large $R_0$, since $\epsilon_{\times}$ goes to zero at large scales, or by combining measurements at different scales \citep{Paranjape2013a,Castorina16a} {\color{black} and solving the resulting set of equations. We use this latter approach to estimate $b_{10}$ and $b_{20}$ and the legends in the figures (for instance \fig{fig-b1M}) specify the scales used. The corresponding equation for $b_2$ has 3 unknown variables and hence requires as many smoothing scales,
\be
\label{eq:b2real}
b_2 = b_{20}+ 2\epsilon_{\times} b_{21}+ \epsilon_{\times}^2 b_{22}
\ee
}
Our first goal is to compare estimates of $b_{10}$ and $b_{20}$ from real and Fourier space. 
We find that both approaches give very similar results. In \fig{fig-b1M} we compare the values of linear and quadratic bias estimated from real space (squares and triangle) and Fourier space (circles). The agreement between the two is very good over the whole mass range we probe and it is very well described by $\text{ESP}\tau$. 
It is important to point out what would have happened if we had not include shear in our bias estimator in Fourier space. This is shown in \fig{fig-b2nos2}, in which the estimates of $b_2$ in Fourier space without taking shear into account are clearly incompatible with the real space results (and with the theoretical model).

We conclude the discussion of real space bias measurements by generalizing the estimator in \eq{eq:bn-real} to non local bias parameters. It is easy to show that the shear field, $s^2$, is $\chi^2$-distributed with five degrees of freedom \citep{BondMyers96,Sheth2001, Sheth2013, Castorina16a}, which implies that the orthogonal polynomials to use in the estimator are Laguerre polynomials ($L_g$),
\begin{equation}
\label{eq:c2-real}
\hat{\tilde c}_{2} = -\frac{1}{r^2}\frac{1}{N}\sum_{i=0}^N L_{g1}^{3/2} \left(\frac{5s_{0,i}^2}{\langle 2 s_0^2 \rangle}\right)
\end{equation}
where $r = S_{\times}/\sqrt{SS_0}$, $s_0$ is the large scale ($R_0$) shear and $\hat{\tilde c}_{2}$ is the corresponding non-local bias \citep{Sheth2013, Castorina16, Castorina16a}. {\color{black} However, unlike our definition of non-local bias in \eq{eq:real-bias}, this estimator assumes that the bias expansion is written in terms of orthogonal polynomials,
\be
1+ \delta_h(M) = 1 + c_1 H_1(\nu) + \frac{c_2}{2} H_2 (\nu) + \tilde{c}_2
L_{g1}^{3/2}\left(\frac{5s_0^2}{\langle 2 s_0^2 \rangle}\right) 
\ee
which are properly normalized and ensure by construction that $\langle \delta_h \rangle =0$. 
Thus, for comparison, our parameter $b_{s^2}$ in Eq. \ref{eq:real-bias}, is related to $\tilde{c}_2$ (upon noting that $\langle s_0^2 \rangle = \frac{2}{3}\sigma^2$, measured on the halo-scale),
\be
b_{s^2} = \frac{15}{4} \frac{\tilde{c}_2}{\sigma^2}\ = \frac{15}{4} c_2\
\ee
}
and from now on, to reduce the dynamic range in the figures, we will only plot $c_2 = \tilde{c}_2/\sigma^2$ and convert our Fourier and PBS estimators to this (we will make use of the words $c_2$, and $b_{s^2}$ interchangeably). 

\fig{fig-b1M}, bottom panel,  shows the large scale values of the $b_{s^2}$ estimated from real and Fourier space. The two agree fairly well, however the real space measurements have still significant errorbars that do not allow us to assert the significance of the excursions of $b_{s^2}$ above zero we see in Fourier Space. {\color{black} As already discussed in \cite{Castorina16a}, real space measurements are mostly affected by stochasticity in the measured values of the shear at large scales, which in combination with $r$ {\color{black}(defined below \eq{eq:c2-real})} being a very small number, make the variance of the estimator very large and increasing with increasing large scale smoothing radius $R_0$. Hence we show the measurements for radius $R_0$ = 20 Mpc/h but for the heaviest halos ($M \sim 10^{15} M_\odot/h$), this is close to their Lagrangian radius and possibly causes the disagreement with other estimates for the heaviest halos in \fig{fig-b1M}.} We plan to come back to this issue, with better measurements, in a future work.
The analytic prediction for the tidal bias is shown as solid line in \fig{fig-b1M}. The agreement with the measurement in the N-body is poorer that for $b_{1(2)}$ and it indicates that, although, the model in \cite{Castorina16} is capable of capturing the gross features of shear into the bias expansions more work is needed to properly understand the origin of the non local bias parameters. 

\section{PBS estimator}
\label{Sec:bPBS}
The PBS definition of halo bias says that bias parameters are the $n$-th order response of the halo population to the presence of large scale fluctuations. For the case of isotropic response, one can exploit the well known equivalence between a infinite wavelength spherical perturbation in a flat Friedmann-Roberston-Walker (FRW) background and a closed FRW Universe. This technique, known as Separate Universe, allows to measure isotropic bias parameters in N-body simulations with basically no cosmic variance  \citep{McDonald2003, Sirko2005, Wagner2015, Lazeyras2016a,Li2016,Baldauf2016}.
The drawback of this method is that it does not apply to tidal bias, where one would need N-body simulations with anisotropic expansion along the different axis, since the introduction of shear breaks isotropy of space. 

\begin{figure}
\includegraphics[width=\columnwidth]{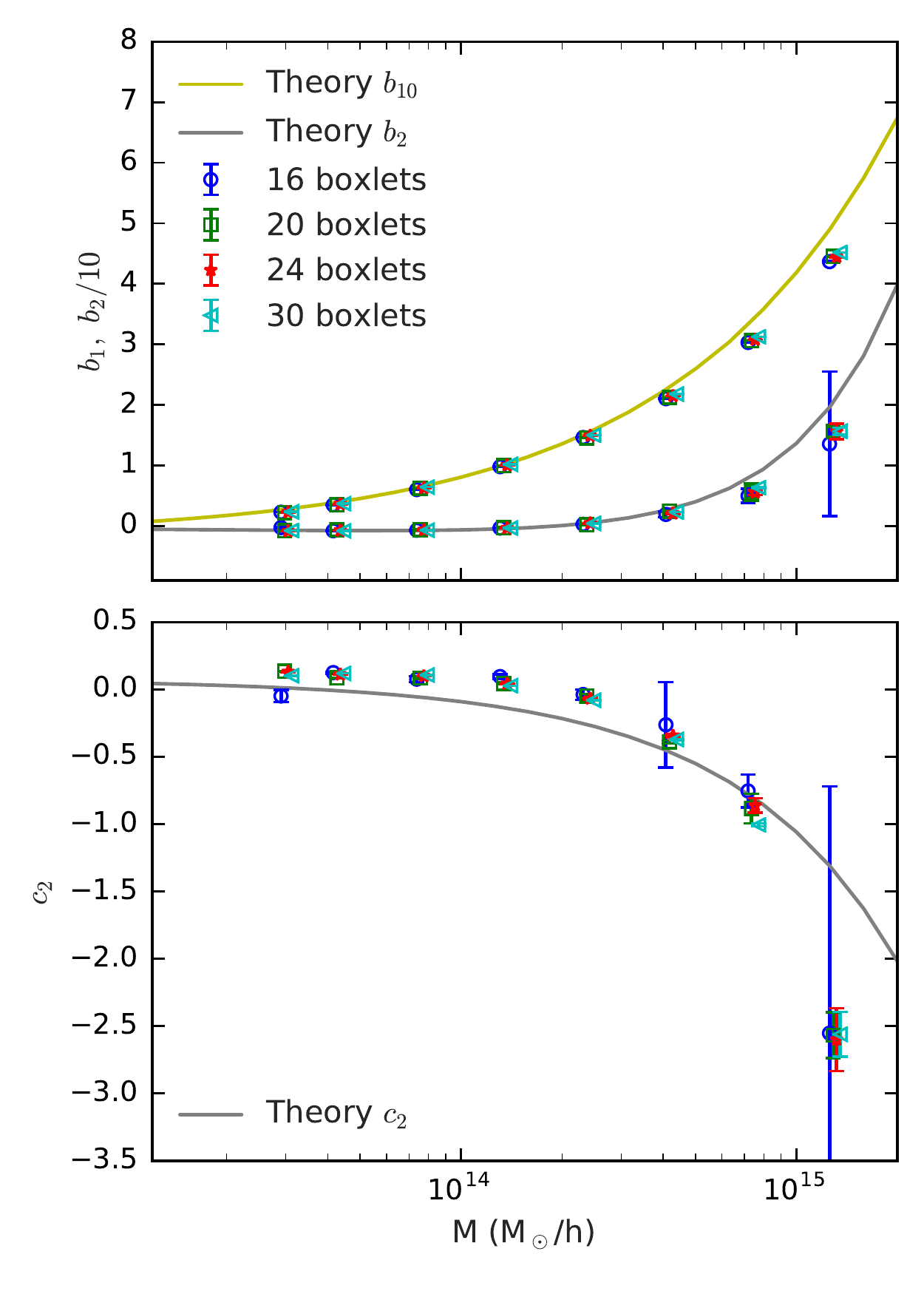}
\vspace*{-5mm}
\caption{{\bf PBS bias measurements from different boxlet sizes}: As explained in the text, the size of error bars increase with increasing boxlet sizes (decreasing number of boxlets). Markers are shifted along x-direction with decreasing numbers of boxlets for clarity.}
\label{fig-sep_univ_boxlets}
\end{figure}

In Lagrangian space, where the density field is linear, one can take advantage of the PBS to measure \emph{any} bias parameters in a much simpler way. 
Consider a relatively large box (of size $L$) of a N-body simulation. {\color{black} We chop this box into several smaller boxes of linear size $l$ in the initial conditions and call them boxlets.} Each boxlets will have its own value of density field, drawn from a Gaussian distribution smoothed at the scale of the boxlet. In order to suppress scale dependent terms one must ensure that $R/l \ll 1$, where $R$ is the typical size of halos. Differently from the standard Separate Universe approach, each boxlets will also have its own value of the shear field, drawn from a $\chi^2$-distribution. 

{\color{black}In the simulations, to estimate the density associated with every boxlet, we smooth the field with a tophat window of the scale of the boxlet, $R_b$ (this scale is computed by matching the volume of a sphere of radius $R_b$ and a cube of the boxlet side $l$). The value of this smoothed density field at the center of the boxlet, $\delta_{i, l}$, is the associated large scale density of the boxlet $i$. From this smoothed density field then, we estimate the shear field using \eq{eq:s2}. The value of this shear field at the center of every boxlet is the associated large scale shear value ($s_{i,l}^2$). We have verified that using the value of these fields at the center of the boxlet is equivalent to computing the mean from all the grid cells belonging to each boxlet, which we expect to be the case since the we are in the linear regime and the variance is small owing to large smoothing scale.}

To evaluate the halo overdensity in every boxlet, we measure the mean number of halos in the box ($\bar{N}$) and in every boxlet ($N_i$), and then compute the relative excess in halo number density as 
\begin{equation}
\hat{\delta}_{h_i} = \frac{N_i/ l^3 - \bar{N}/L^3}{\bar{N}/L^3}\;.
\end{equation}
Bias parameters can then easily be obtained by fitting the above quantity to the measured density and shear in each boxlet
\be
\label{eq:PBS}
\hat{\delta}_{h_i} = b_{10} [\delta_{i,l}] +\frac{1}{2}b_{20} {[\delta_{i,l}]}^2 + b_{s^2}[s_{i,l}]^2 + ...
\ee
{\color{black} Here we have modified the notation as compared to \eq{eq:real-bias} and in accordance with \cite{BiasRev}. This is to make it explicit that these (and in other moment based methods) bias parameters are defined with respect to an underlying density field smoothed at large scales ($l \equiv R_b$ in our case) and hence should not be compared directly against other methods which do not make reference to any smoothing scale, such as our fourier-space based method. We use the mapping described in \cite{BiasRev} to compare the two bias estimates, and we refer the reader to their discussion of moments based estimates for detailed discussion. Briefly, to first order, the mapping leaves linear bias unchanged while linearly mixes the two second order bias estimates $b_2$ and $b_{s^2}$ through a matrix whose coefficients depend on the smoothing scale $R_b$. We find that for scales as large as ours, this mixing does not lead to any significant change in either bias parameters.}

{\color{black} While this method has several advantages over standard techniques it has its own drawbacks. Since the boxlets are large, $\mathcal{O}(100) \,\Mpc$, the value of the density and shear are small and the scatter among different boxlets large. This makes difficult to fit for bias parameters and it yields slightly larger error bars than the standard Separate Universe technique. This can in principle be alleviated by decreasing the boxlet size and hence increasing the magnitude of super sample density and shear. However, this leads to high mass halos being poorly sampled in the small volume of the boxlets, and being affected by non-poissonian shot noise, which is hard to estimate. Furthermore, as previously mentioned, the boxlets need to be sufficiently larger than the Lagrangian halo scale to suppress scale dependent terms.} 

\fig{fig-sep_univ_boxlets} shows bias coefficients estimated using \eq{eq:PBS} for different chopping of the $L=3000 \,\Mpc$ box, $L/l = 30,\,24,\,\,20,\,16$. For various size of the boxlets the measured bias parameters agree among themselves, and we also report clear evidence of tidal bias with this novel estimator. As discussed, the size of error bars decreases with decreasing the size of boxlets (increasing number) due to better constraining power. However, to safe-guard against possible errors described above, we will henceforth quote values for $L/l = 20$ in the remainder of this work. 

\begin{figure}
\includegraphics[width=\columnwidth]{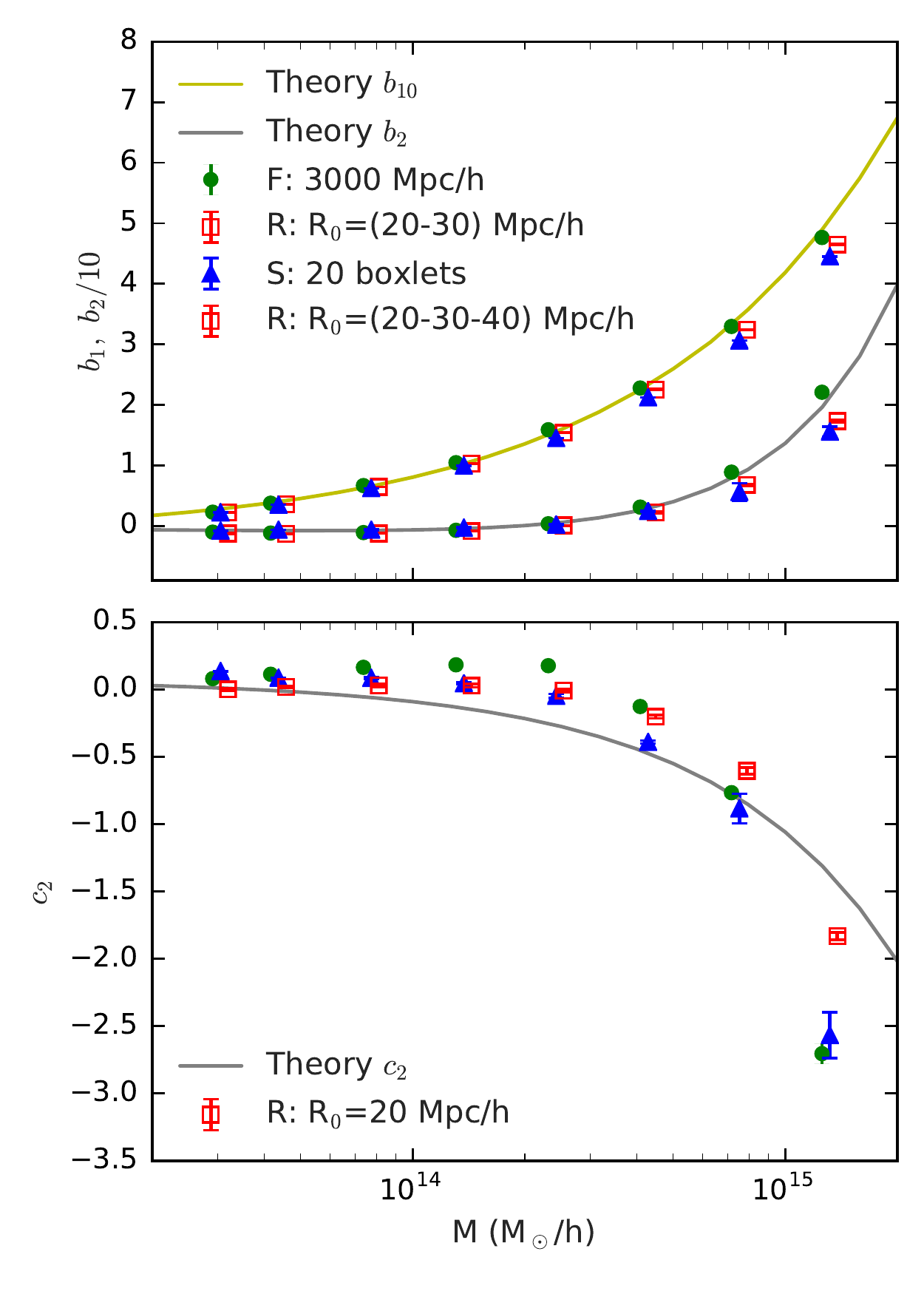}
\vspace*{-5mm}
\caption{{\bf PBS $b_1$, $b_2$ and $b_{s^2}$ as function of mass}: comparison with ESPTau theory, fourier space and real space measurements. Measurements are shown only for $3\ Gpc/h$ box and PBS measurements for 20 boxlets. The PBS and real space points have been shifted along x-direction for clarity.}
\label{fig-sep_univ_b1M}
\end{figure}

The concordance between the three different estimators used in this paper, Fourier Space, real space and PBS, is shown in \fig{fig-sep_univ_b1M} and it is one of our main results. We stress that, similarly to what happened in Fourier Space, in fitting \eq{eq:PBS}, the presence non local bias was crucial to obtain the agreement for local linear and quadratic bias between different estimators, as well as the theory. While we do find convincing evidence of tidal bias in all three measurements and there is general agreement in the values, a \% level calibration of non local bias would require more simulations and better control of the errors, and it goes beyond the scope of this paper. We plan to return to it in a future work.

\section{Relations among the bias parameters}
\label{Sec:rel}
Relations among the bias coefficients are of great value for cosmological analyses as they make possible to reduce the number of free parameters. In this section we study three kinds of such relations: between the the scale dependent and the scale independent terms of linear bias, the general universality of these parameters with redshift and empirical relations between $b_2$ or $b_{s^2}$ and $b_1$ that universality allows one to fit for.

\subsection{Consistency relation for linear bias}	
\begin{figure}
 \includegraphics[width=\columnwidth]{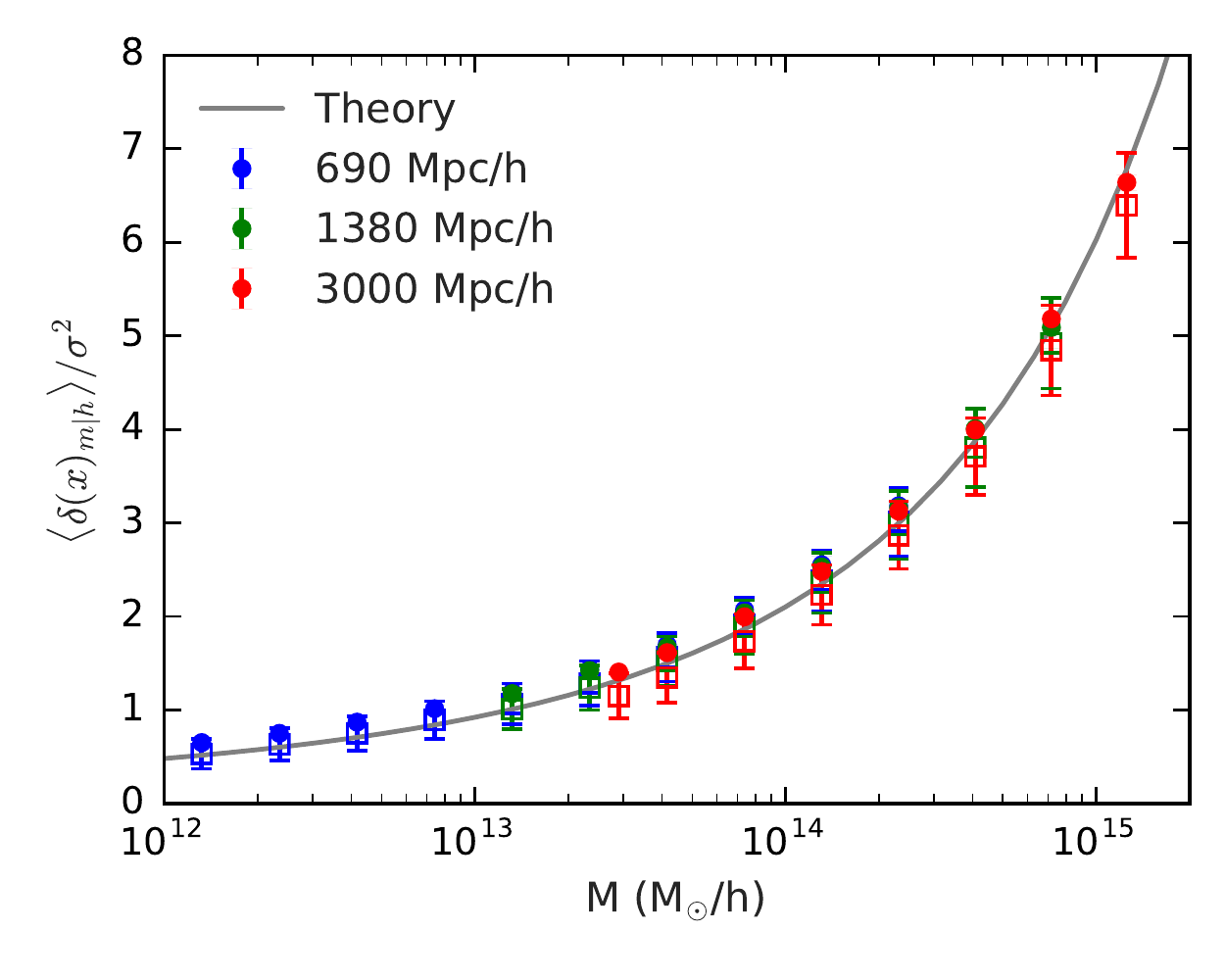}
 \caption{{\bf Consistency Relation between $b_{10}$ and $b_{11}$:} \eq{eq:consrel} relates the linear bias to the mean dark matter overdensity at the position of the halos. For Spherical collapse, this mean $\langle \delta(x)_{m|h} \rangle$ = 1.686, but its a function of mass in Ellipsoidal collapse. The square markers (shifted along x-direction for clarity) are the direct measurements from the simulations while solid dots are prediction from bias parameters estimated from fourier space correlations in previous sections from those catalogs.}
\label{fig-consistency}
\end{figure}
In the Lagrangian picture, properties of the (proto-)halos profiles are closely related to the bias parameters. From the very definition of halo mass and halo bias in \eq{eq:b1real}, it is easy to see that at the scale of the halo $\epsilon_\times(R) = 1$, and therefore \citep{Musso2012,Paranjape2013,Paranjape2013a,Castorina16, Castorina16a}
\be
\label{eq:consrel}
b_{10}+b_{11} = \frac{\langle \delta_{1\times} \rangle}{\sigma^2} 
\ee
where $\langle \delta_{1\times} \rangle$ is the mean dark matter density inside halos of a given mass and $\sigma^2$ is the variance on the scale of halos. If halos were perfect spheres, then $\langle \delta_{1\times} \rangle = \delta_c=1.686$. Measurements of this quantity have been presented in \cite{Sheth2001, Dalal2008, Robertson2009,Ludlow2013,Despali2016,Castorina16}.
The above equation says something very non-trivial about bias coefficients: one can measure the value of bias parameters at large scales and learn something about the density of halos at a much smaller scale, the halo scale. 
It is also important to note that \eq{eq:consrel} remains valid even in the presence of assembly bias, as it is a direct consequence of the definition of mass \citep{Castorina16a}.
We have seen in \fig{fig-b11_mass} and  \fig{fig-b1M} that different estimates of linear bias agree with each other and with a theoretical model. We thus expect the consistency relation to hold in the data and to be well described by the $\text{ESP}\tau$ model. This is shown in \fig{fig-consistency}, where one can see how the \eq{eq:consrel} is verified at the level of the data, blue and red points with error bars, and how the theory actually does a good job in predicting it.  
In real space it is much harder to check the consistency relation for linear bias, see \cite{Castorina16}, while a similar analysis in Fourier space can be found in \cite{Chan16}.
Since the $k^2$ term in linear bias, $b_{11}$, is highly degenerate with the leading order counter term $\alpha$, in the Effective Field Theory (EFT) of Lagrangian perturbation theory \citep{Vlah2016}, one could argue that  it should be possible to just fit for $b_{11}$ in a EFT approach, and then use the consistency relation to put a prior on the value of scale dependent bias.
Similar relation should hold between higher order bias parameters and higher moments of the density field at the halo scale.

\subsection{Universality of Bias Relation with redshift}
In the study of halo abundances and clustering, universality means that the halo mass function and halo bias are unambiguously determined by a single function of cosmology and redshift, the variance of the linear density field, $\sigma(z)$. The existence of universal relations is important, since it means we have a proxy for how the bias of a population of halos evolves, for instance, with redshift. Violations of the universality of the mass function in N-body simulations with redshift have been reported in \cite{Tinker2008,Crocce2010,Watson2013}, while the Eulerian linear bias remains a universal function \citep{Sheth1999,Tinker2010,Hoffmann2015}
The purpose of this section is to test universality of linear and non linear local bias, as well as the novel $b_s^2$, with redshift. 

\begin{figure}
 \includegraphics[width=\columnwidth]{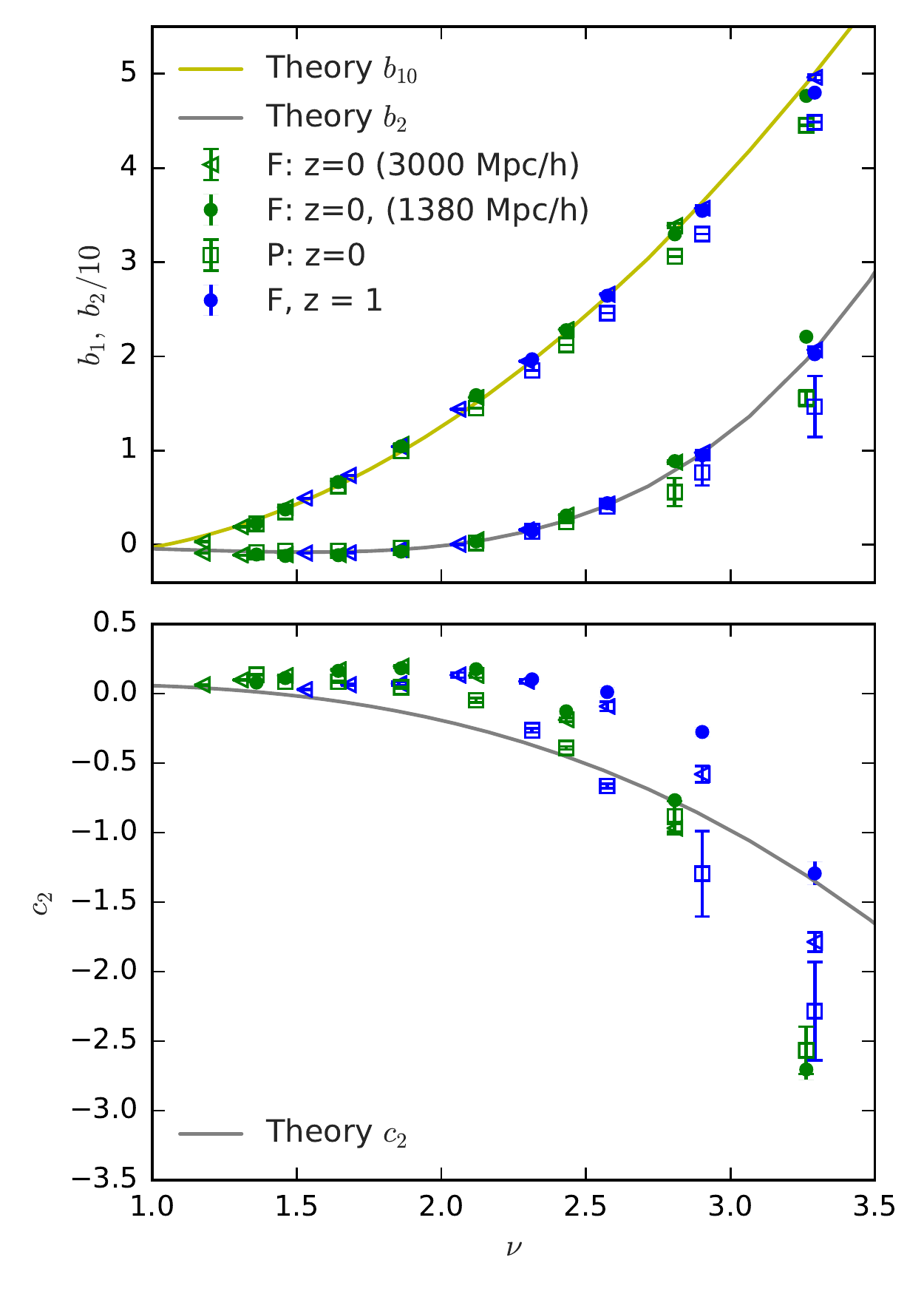}
\vspace*{-5mm} 
\caption{{\bf Universality of bias parameters:} agreement between our fourier space and PBS estimator across the redshifts, blue points are from $z=0$ and green points are $z=1$. PBS measurements are only for $3\ Gpc/h$ box. The analytic prediction is assumed to be redshift independent.}
\label{fig-universal}
\end{figure}

{\color{black} The best way to test universality is to plot quantities estimated at different redshift as function of the peak height $\nu = 1.686/\sigma(z)$, which takes out the redshift and cosmology dependence of the bias parameters. If universality holds, bias parameters across redshifts will align for the same value of $\nu$. Thus, we repeated our analysis of the previous sections at $z=1$, and show the bias coefficients from both redshifts in \fig{fig-universal}. The values of $b_1$ and $b_2$, in blue at $z=0$ and in green at $z=1$ in the top panel, align very nicely as a function of $\nu$ over all the mass range probed by our simulations. For tidal bias, the PBS measurements (empty squares) do show that the bias is approximately universal, whereas the Fourier space measurements  show some non-universality at high mass halos. This disagreement between the two estimates can possibly be due to poor fits, as discussed at the end of Section \ref{Sec:bF}. The continuous line is the $\text{ESP}\tau$ theory from $z=0$, for which we have thus implicitly assumed redshift independence, which consistently predicts both $b_1$ and $b_2$. It is worth pointing out that there is no \textit{a priori} reason for the bias parameters to be universal. In the framework of Excursion Sets/Peaks theory  this would imply the two parameters in \eqref{eq:ESPtau}, $\delta_c$ and $\beta$, are redshift independent. While it is well known that spherical collapse yields universal mass function and bias, the value of $\beta$, an effective way to measure the relative importance of anisotropies in the collapse process, can have arbitrary redshift evolution. Since the value of $c_2$ is very sensitive to the value of $\beta$, we expect tidal bias to be a better indicator of any violation of universality.} 
A detailed study of the importance of the shear for halo formation as a function of redshift goes beyond the scope of this work, and we intend to come back to this issue in the future.

\subsection{Relation between higher order bias parameters}
\begin{figure}
\includegraphics[width=\columnwidth]{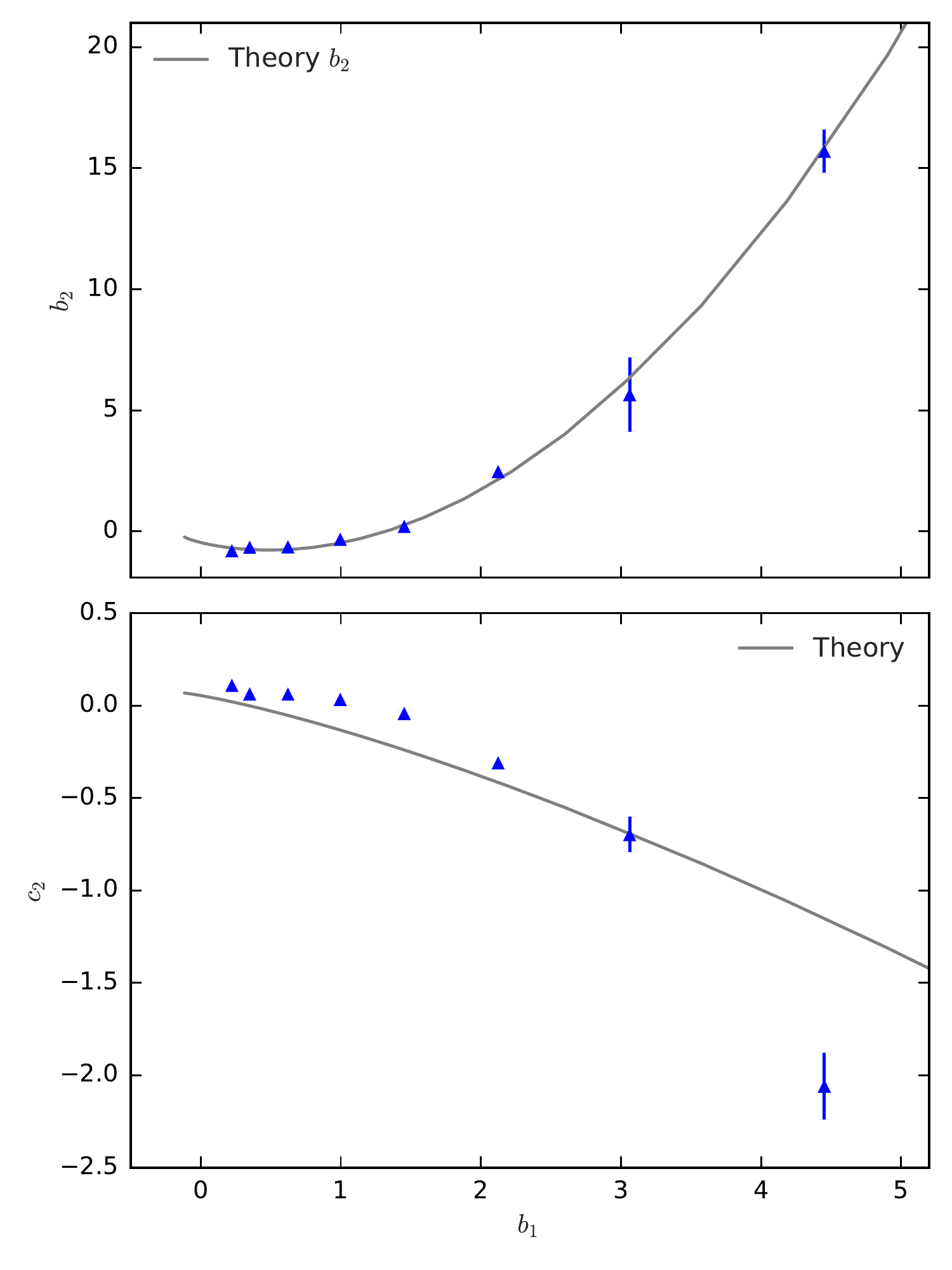}
\vspace*{-5mm}
\captionsetup{font={small}}
\caption{{\bf Relation between bias parameters:} Empirical relations between $b_2 - b_1$ (upper panel) and $b_{s^2} - b_1$ (lower panel) shown only for PBS estimator since tidal bias is most universal for these. The solid lines are using bias values predicted by $\text{ESP}\tau$ theory. These relations, co-evolved to Eulerian space help to reduce dimensionality of bias parameter space.}
\label{fig-relation}
\end{figure}

Another way to restrict the dimensionality of the bias parameters space is to find relations between linear bias and higher order bias parameters. While such relations can be determined at each redshift, if the bias parameters are universal, then one could just fit for $b_{2}(b_1)$ and $b_{s^2}(b_1)$. 
In an Eulerian framework this has been discussed a lot in the literature, see recent work by  \cite{Hoffmann2015,Hoffmann2016, Lazeyras2016a}.
However it is important to realize that if such relations exist, they do because they hold in Lagrangian Space.
We must also stress another word of caution: the nature of these relations is empirical, not fundamental, and it is, for instance, violated by assembly bias effects. 
Figure \ref{fig-relation} shows $b_2$ as a function of $b_1$, upper panel, and $b_{s^2}$ as a function of $b_1$, lower panel. Since the PBS estimates of tidal bias are universal, we show the relations only for these data.
The measured values smoothly align, indicating that the relation could be fitted by some simple functional form. As expected from results in the previous sections, the theory is able to predict very well the relation between non linear and linear bias, but it is not as accurate for the tidal bias. 

\begin{figure}
\includegraphics[width=\columnwidth]{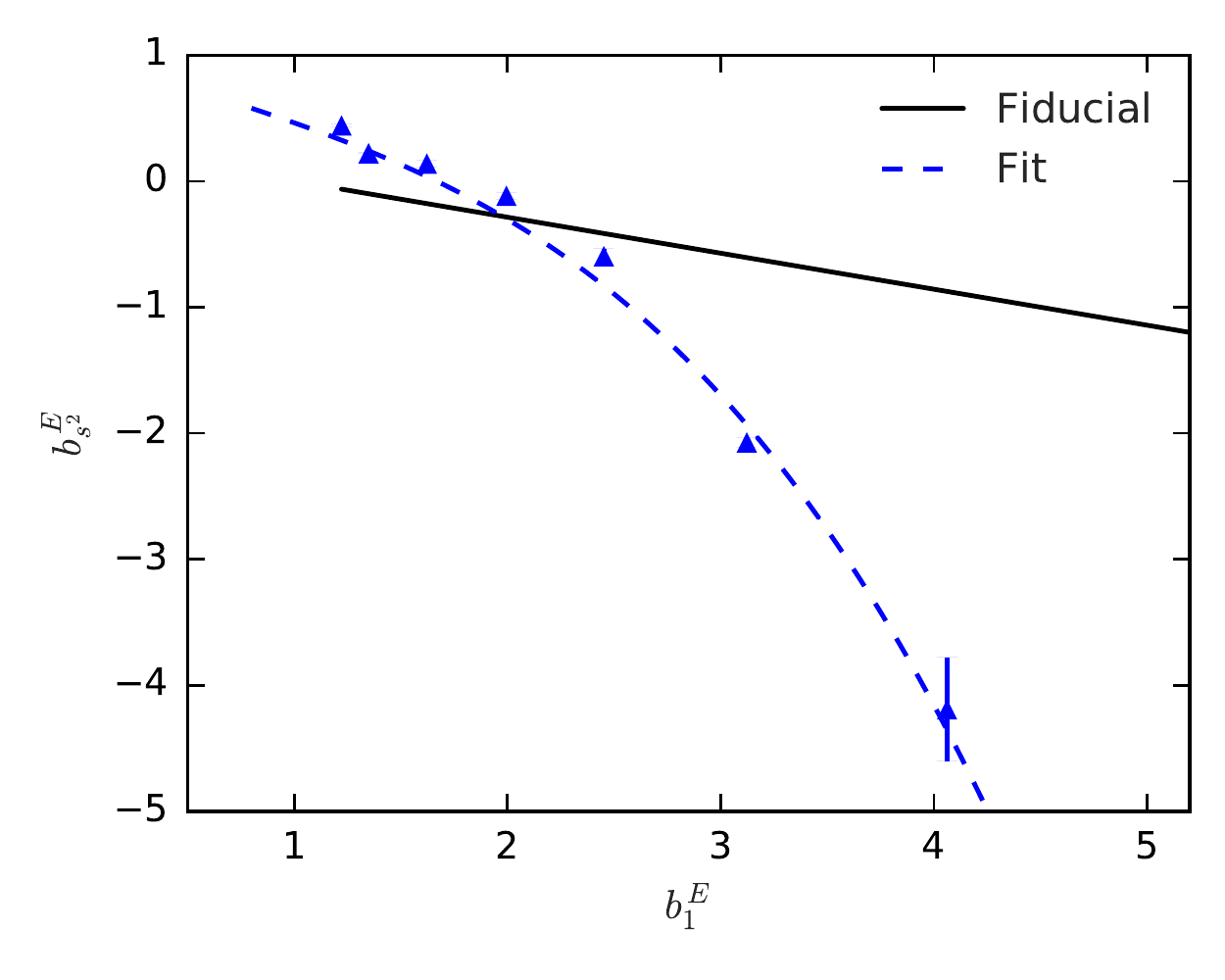}
\vspace*{-1mm}
\captionsetup{font={small}}
\caption{{\bf Eulerian Non-Local bias:} Comparison of the evolution relation for shear bias in Eulerian space with (\eq{eq:bs2_eul}; markers) and without (\eq{eq-shear_coev}; solid black line) the presence of Lagrangian shear bias as measured by PBS estimator. The numerical best fit (\eq{eq:bs2_fit}) to the measurement points is in dashed blue.}
\label{fig-bs2Eul}
\end{figure}

We want now to go back to the relation between Lagrangian and Eulerian bias parameters  as discussed in the Introduction. If $b_{s^2}$ was zero, gravitational evolution makes a unique prediction for the Eulerian tidal bias,  $b_{s^2}^E$, \citep{Chan2012,Baldauf2012} 
\be
b_{s^2}^E = -\frac{2}{7}b_1 = -\frac{2}{7}(b_1^E-1) 
\label{eq-shear_coev}
\ee
where we have used the well known relation $b_1^E = 1+b_1$. 
As we have measured a non-zero shear bias in Lagrangian space, the above equation needs to be modified \citep{Sheth2013},
\begin{equation}
\label{eq:bs2_eul}
b_{s^2}^E = b_{s^2} - \frac{2}{7}(b_1^E - 1)
\end{equation}
Figure \ref{fig-bs2Eul} shows the relation between linear and tidal bias in Eulerian space, using the measurement from the PBS method. The assumption of no Lagrangian shear bias, black line, is far from the measurements at both low and high value of $b_1^E$. Thus, in the presence of this Lagrangian shear bias, using the relation of \eq{eq-shear_coev} such as done in analysis of \cite{Gil-Marin2015, Beutler2016,Sanchez2017} can lead to possible systematic errors in the analysis. Since the analytic model does not yield an accurate fit to the data, for practical purposes we provide a numerical fit to the PBS measurements:
\begin{align}
\label{eq:bs2_fit}
b_{s^2}^E (PBS)&= 1.03 -0.615 b_1^E + 0.118(b_1^E)^2 - 0.072 (b_1^E)^3 
\end{align}
{\color{black}Thus we expect $b_{s^2}^E \sim 0$ at $b_1^E \sim 1.7$ which corresponds to halos of mass $M \sim 10^{14}$ M$_\odot$/h at $z=0$ and $M \sim 7 \times 10^{12}$ M$_\odot$/h at $z=1$.}

\section{Stochasticity}
\label{Sec:stoch}
Following \cite{Seljak2004,Hamaus2010,Baldauf2013}, for any two fields $X$ and $Y$, where the field $Y$ is supposed to model field $X$, we measure the error in modeling in terms of the stochasticity, $S_{XY}$,  defined as 
\begin{align}
\label{eq:stoch}
S_{XY} = P_{XX} - \frac{P_{XY}^2}{P_{YY}} \,
\end{align}
For our purpose here, we are interested in examining how well we reconstruct the halo field with the estimated bias parameters. {\color{black} Thus $X$ is the halo overdensity field and $Y$ is the corresponding field modeling the halo field and given by the right hand side of the \eq{eq:real-bias}. For the modeling field ($Y$), we successively add the higher order bias parameters, starting from $b_1$ to $b_2$ and $b_{s^2}$ so that we can identify the improvement contributed by every successive parameter. Then, $P_{XX}$ is the halo power spectrum ($P_{hh}$), while $P_{YY}$ is the auto-power spectra for the modeling (bias) field ($P_{bb}$, where $b$ will be the highest order (in terms of complexity) bias used).} The explicit expression for $P_{bb}$ as well as cross-spectra $P_{XY}$ in terms of different bias parameters is
\begin{align}
\label{eq:power_stoch}
P_{XY} & \equiv P_{hb} = b_1 P_{h \delta} + \frac{b_2}{2} P_{h \delta^2} + b_{s^2}P_{hs^2} \\
P_{YY} & \equiv  P_{bb} = b_1^2 P_{\delta \delta} + \frac{b_2^2}{4} P_{\delta^2 \delta^2} + b_{s^2}^2P_{s^2s^2} + b_2 b_{s^2}P_{\delta^2 s^2} \nonumber
\end{align}
If the modeled bias-field reconstructs the halo field perfectly, all the three power spectra in the definition of stochasticity should be the same resulting in $S_{hb}=0$. However since the number of halos, ${n}$, is finite, the halo auto spectrum contains Poisson shot noise, $\frac{L^3}{n}$ (where $L$ is the box size), which is not captured by the continuous bias field.

\begin{figure}
\includegraphics[width=\columnwidth]{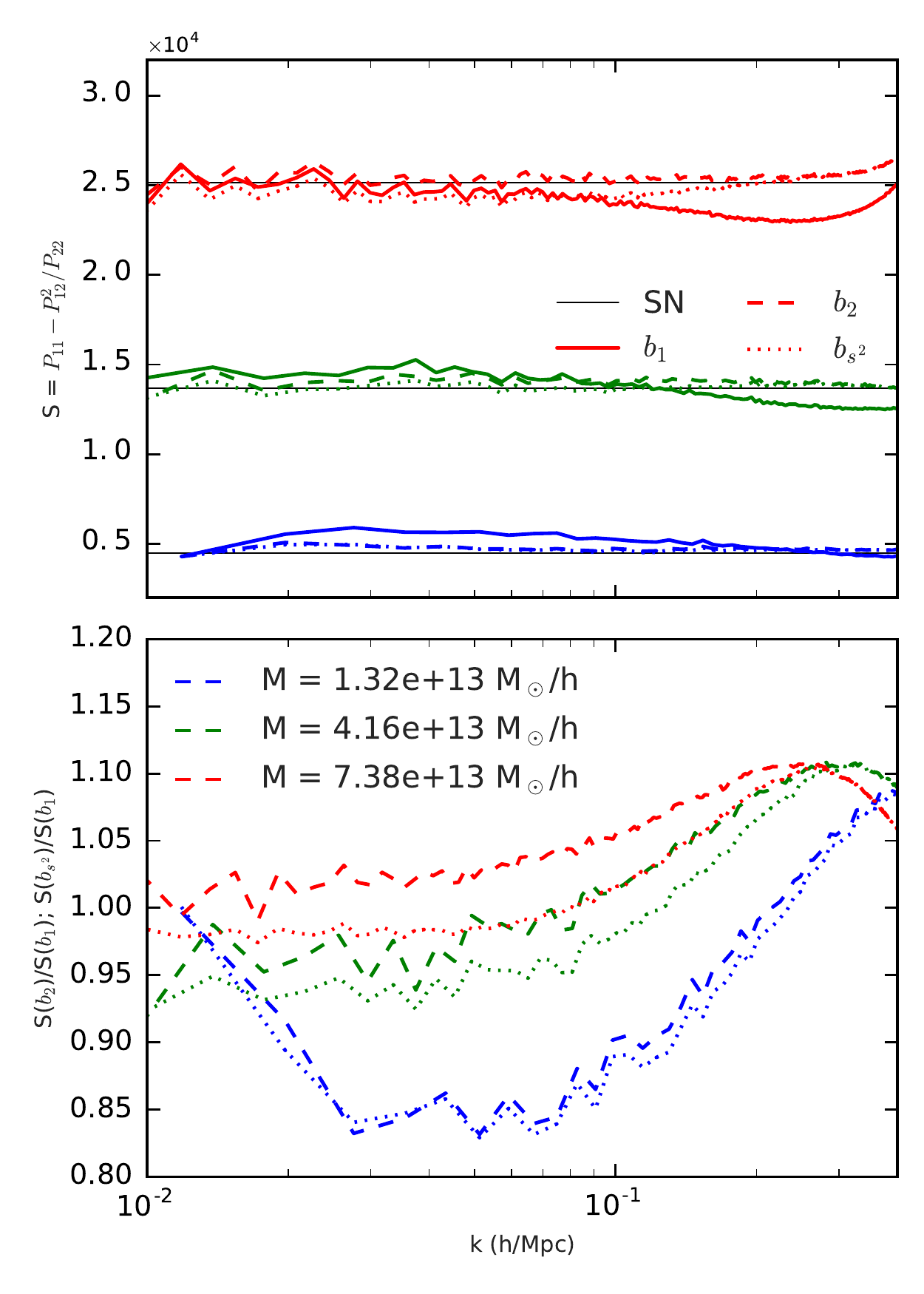}
\vspace*{-5mm}
\caption{{\bf Stochasticity of Halos:} The stochasticity in the upper panel is calculated using \eq{eq:stoch}. The solid lines, dashed and dotted lines respectively correspond to measurements on successively including $b_1$, $b_2$ and $b_{s^2}$ in \eq{eq:power_stoch}. For comparison, the black lines are shot noise for that mass bin. The lower panel shows the ratio of dashed and dotted lines of upper panel with the solid lines to emphasize that more complex bias models reduce stochasticity.}
\label{fig-stochasticity}
\end{figure}

In \fig{fig-stochasticity}, we show  measurement of stochasticity in for three different mass bins, one from each of the three simulation boxes. For each mass, we show how the stochasticity changes as we include higher order bias parameters over simple linear bias $b_1$ in \eq{eq:power_stoch}. {\color{black}To estimate stochasticity, we first average the various power spectra for all 5 simulations of a given size to reduce noise and cosmic variance in individual measurements on large scales. These mean spectra are then used with the constant (large scale) Fourier space bias estimates in \eq{eq:stoch}.}

Overall, on large scales, the stochasticity is close to its Poisson shot noise value for intermediate and low mass bin, while its somewhat lower for heavier halos (\cite{Baldauf2013} explains this as exclusion effects). However, especially for the low mass halos, simply using the linear bias for halo field does leave significant scale dependence in the stochasticity which is improved upon by including higher order bias parameters (upper panel). This is useful since this residual can still be modeled with a scalar, if not necessarily poisson, shot noise. In addition, including $b_2$ and especially $b_{s^2}$ does assist in reducing stochasticity further over linear bias models (lower panel). 

\section{Discussion and Conclusions}
\label{Sec:conclude}

Understanding the relation between the galaxy or halo field and the underlying dark matter distribution, is crucial in order to extract cosmological information from the LSS.

In this paper we have focused our attention to halo bias defined in Lagrangian Space. 
We employed three different estimators of bias parameters- Fourier space correlations, real space correlations and PBS estimator.  The three approaches are quite different and thus sensitive to different systematic effects, nevertheless their agreement in terms of best-fit bias parameters as a function of halos mass is quite remarkable (Figure \ref{fig-sep_univ_b1M}). 

We have shown a convincing evidence for shear bias in Lagrangian space with all three methods (Figures \ref{fig:bias-kgrid}, \ref{fig-b1M} and \ref{fig-sep_univ_b1M}). Indeed, including the tidal bias was crucial to obtain the consensus of quadratic local bias of all the three estimators (Figure \ref{fig-b2nos2}). We have shown that the model presented in \cite{Castorina16} is able to predict well $b_1$ and $b_2$, but it misses a good description of tidal bias. 

The Fourier space estimator have also enabled us to present the scale dependence of quadratic and shear bias (Figure \ref{fig:bias-kgrid}), and to show that it is similar to the one of linear bias. For linear bias, the Fourier space method allowed us to check that the scale dependent piece (Figure \ref{fig-b11_mass}) is very well predicted by theory. This has implications for the problem of what is the right window to use when defining halos in Lagrangian space. We also successfully demonstrated, in N-body simulations, the validity of a consistency relation for linear bias coefficients, \eq{eq:consrel} and \fig{fig-consistency}, that are fundamental to the bias parameters. This opens the door to reduce the number of independent bias parameters. 

We have then used the PBS argument to generalize previous results on bias with respect to density, to the case of bias with respect to the shear field.
We were therefore able to measure non-local bias as the response of the halo number density to the presence of long wavelength tidal field (Figure \ref{fig-sep_univ_boxlets}). In the appendix \ref{Sec:Appendix-A}, we also show that this method can be extended to third order bias parameter $b_3$ (Figure \ref{fig-sep_univ_b3}). However care needs to be taken while choosing the size of boxlets due to the dichotomy between better constraining power for small boxlets but at the possible cost of residual scale dependence and non-Poissonian shot noise.

Relations among linear bias and quadratic bias parameters have also been discussed, with the general finding that those relations exist and can be easily fitted for (Figure \ref{fig-relation}). Along the way we also showed that all these bias parameters are to a very good approximation a universal function of redshift (Figure \ref{fig-universal}).

We note that while the evidence of non-zero tidal bias seems convincing and there is general agreement between different estimators, Fourier space shear estimates make larger excursions above zero than real space or PBS estimates (Figure \ref{fig-sep_univ_b1M}), and also seem to exhibit some violations of universality for high mass halos. While there is no a priori reason to expect density and shear bias coefficients to exhibit the same level of universality, we discussed in Section \ref{Sec:bF} whereas the possible reason could be poor fits due to lack of correct halo window as well as complicated scale dependence of tidal bias, coupled with degeneracy of $\delta^2$ and $s^2$ on very large scales.  This is an important point which we wish to investigate in future work, with better measurements.

As we have shown, the analytic model in \cite{Castorina16} is not capable to reproduce our measurements of the tidal bias, Figure \ref{fig-bs2Eul}, therefore we provide a numerical fit to the relation between the Eulerian linear bias, $b_1^E$, and the Eulerian shear bias, $b_{s^2}^E$, in  \eq{eq:bs2_fit}.

Finally, we discuss the stochasticity of the halo field and find that including more complex bias models over simple linear bias does seem to reduce stochasticity, especially its scale dependence (\fig{fig-stochasticity}). This makes these models worth studying since a scale independent stochasticity can be parameterized with a constant shot noise which can be marginalized over in the analysis. 

\section*{Acknowledgments}
We would like to thank Fabian Schmidt for pointing out the caveat in comparing PBS estimates with other methods, as discussed in Section \ref{Sec:bPBS} below \eq{eq:PBS}. EC would like to thank Ravi Sheth, Aseem Paranjape and Martin White for useful discussions. This work is supported by NASA grant NNX15AL17G.

\bibliographystyle{mn2e}
\bibliography{references}

\appendix{}
\section{Third order bias}
\label{Sec:Appendix-A}
In principle, we should be able to extend the above estimates to higher order bias parameters. In this appendix, we provide details on our attempts to do so. For the real space estimates, this involves calculating the mean of third Hermite polynomial at the position of the halos, \eq{eq:bn-real}, which is straightforward. Our attempts to extend the Fourier space estimate along the lines of \eq{eq:b2k-bs2k} return very noisy estimates for $b_3$ on large scales. This is due to $P_{h\delta}\ \& \ P_{h\delta^3}$ as well as $P_{\delta  \delta},\ P_{\delta^3  \delta}\ \& \ P_{\delta^3  \delta^3}$ being degenerate on large scales. Refer to the end of section \ref{Sec:bF} for a similar discussion.\\

\begin{figure}
 \includegraphics[width=\columnwidth]{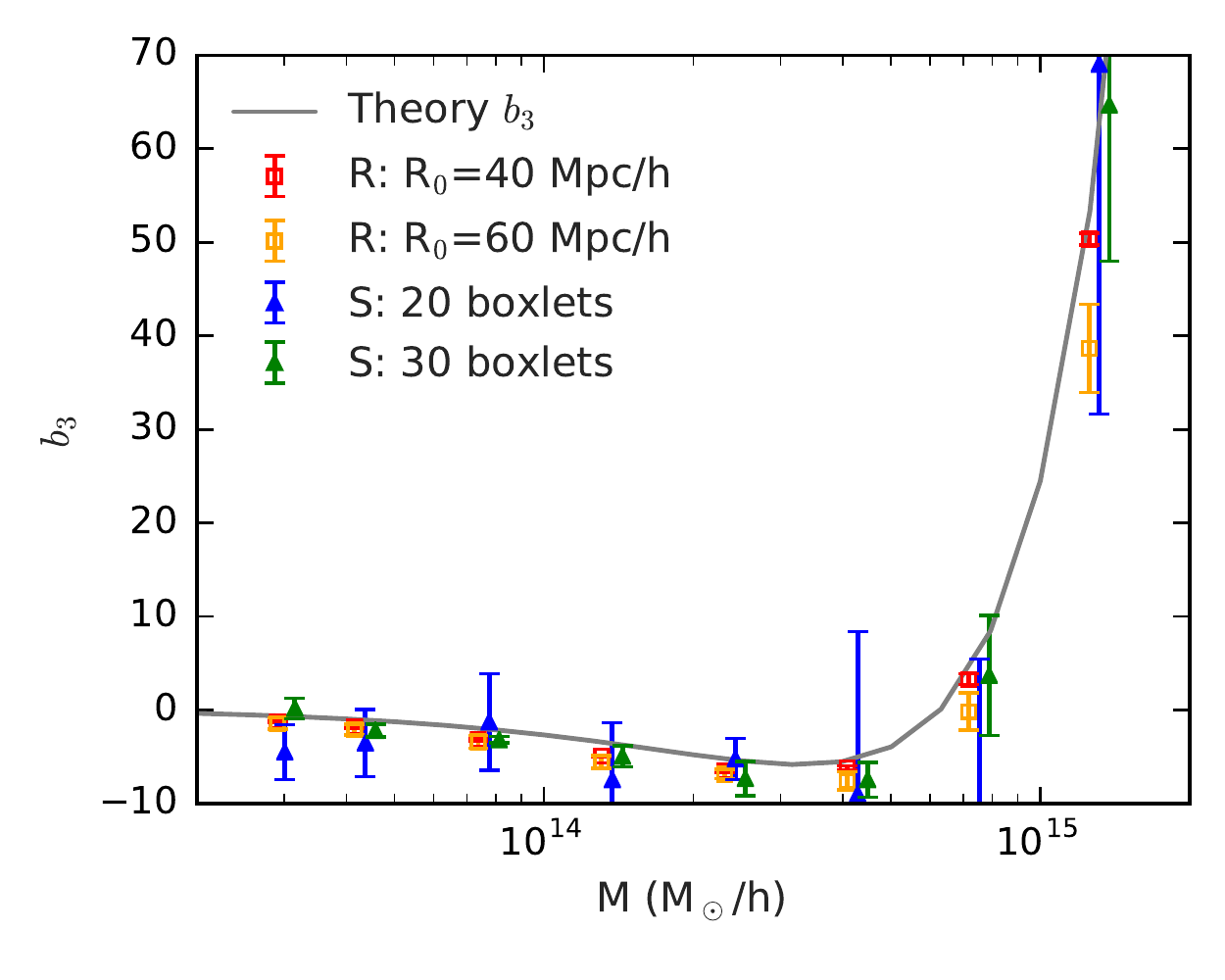}
 \caption{{\bf $b_3(M)$ from real space and PBS}: The trend qualitatively agres with the theory but PBS measurements have significant error bars.}
\label{fig-sep_univ_b3}
\end{figure}

In the PBS case, we can simply extend \eq{eq:PBS} as following and fit for it:
\begin{equation}
\delta_h(\bx) = b_1 \delta(\bx) + \frac{b_2}{2}\delta^2(\bx) + \frac{b_3}{6}\delta^3(\bx) +b_{s^2}s^2(\bx)
\end{equation}
{\color{black} Note that for simplicity, here we ignore the distinction that this method gives bias parameters with respect to the field smoothed on large scale ($l \equiv R_b$) since we found the difference between the two set of parameters to be negligible (see discussion after \eq{eq:PBS}).\\
The results for real space and PBS estimates are show in Figure \ref{fig-sep_univ_b3}, where we are showing the real space results only from 3 Gpc/h box, to compare them with their separate universe counterparts. While the results generally agree, the error bars are quite big in the PBS case, and are larger for 20 boxlet case than 30 boxlets due to lower constraining power of larger boxlets as discussed in Section \ref{Sec:bPBS}. 
}
\label{lastpage}

\end{document}